%
\input harvmac
\def\Title#1#2#3{#3\hfill\break \vskip -0.35in
\rightline{#1}\ifx\answ\bigans\nopagenumbers\pageno0\vskip.2in
\else\pageno1\vskip.2in\fi \centerline{\titlefont #2}\vskip .1in}


\def\ket#1{| #1\rangle}

\def\half{{1\over 2}}

\def\R{\hbox{\rm I \kern-5pt R}}

\font\ticp=cmcsc10
\def\ajou#1&#2(#3){\ \sl#1\bf#2\rm(19#3)}
\def\ul#1{\underline{#1}} 
%
%
\lref\wiesner{S.~Wiesner, \ajou SIGACT News &15 (83) 78.}
\lref\BBeightyfour{C. H. Bennett and G. Brassard, in {\it Proceedings of IEEE
 International Conference on Computers, Systems and Signal Processing} (IEEE,
 New York, 1984), p.~175.} 
\lref\lochauprl{H.-K.~Lo and H.~Chau, \ajou Phys. Rev. Lett. &78 (97) 3410.}
\lref\mayersprl{D.~Mayers, \ajou Phys. Rev. Lett. &78 (97) 3414.}
\lref\mayerstrouble{D.~Mayers, quant-ph/9603015.} 
\lref\lochau{H.-K.~Lo and H.~Chau, \ajou Physica D &120 (98) 177.}
\lref\lo{H.-K.~Lo, \ajou Phys. Rev. A &56 (97) 1154.} 
\lref\blum{M.~Blum, in {\it Proceedings of the 24th IEEE Computer
Conference ( Compcon)}, pp. 133-137f (1982).}
\lref\kilian{J. Kilian, {\it Uses of Randomness in Algorithms and
Protocols } (MIT Press, Cambridge, Massachusetts, 1990), p.68} 
\lref\mayersone{D. Mayers, in {\it Proceedings of the Fourth Workshop on
 Physics and Computation} (New England Complex System Inst., Boston, 1996),
 p.~226.} 
\lref\bcms{G.~Brassard, C.~Cr\'epeau, D.~Mayers and L.~Salvail, 
quant-ph/9806031.}
\lref\akbcrel{A.~Kent, Phys \ajou Phys. Rev. Lett. &83 (99) 1447-1450.}
\lref\akbcreltwo{A.~Kent, ``Secure Classical Bit Commitment over Finite 
Channels'', quant-ph/9906103, submitted to J. Cryptology.}


\Title{\vbox{\baselineskip12pt\hbox{ DAMTP-1998-123}\hbox{quant-ph/9810067}{}}
}{\centerline{Coin Tossing is Strictly Weaker than Bit Commitment}}{~}

\centerline{{\ticp Adrian Kent}}
\vskip.1in
\centerline{\sl Department of Applied Mathematics and
Theoretical Physics,}
\centerline{\sl University of Cambridge,}

\centerline{\sl Silver Street, Cambridge CB3 9EW, U.K.}

\bigskip

\centerline{\bf Abstract}
{We define cryptographic assumptions applicable to
two mistrustful parties who each control two or 
more separate secure sites between which special 
relativity ensures a time lapse in communication.  
We show that, under these assumptions, unconditionally secure 
coin tossing can be carried out by exchanges of 
classical information.
We then show that, under standard cryptographic 
assumptions, coin tossing is strictly weaker than bit commitment. 
That is, no unconditionally secure bit commitment protocol
can be built from a finite number of invocations of a secure coin
tossing black box together with finitely many additional
classical or quantum information exchanges.  
\vfill
Electronic address: apak@damtp.cam.ac.uk}
\eject
The problem of remote coin tossing was introduced into the 
cryptographic literature in a classic 1981 paper by Blum.\refs{\blum}
A coin tossing protocol involve two mistrustful 
separated parties who wish to use an information channel --- 
e.g. a phone line --- to generate a bit in whose
randomness both are confident. 

Coin tossing is a simple cryptographic primitive with many applications 
in more complicated tasks.  To give a well known example, it 
can be used to authenticate
a remote user, say to their bank, as follows. 
The user's $N$ digit passkey is known to both user and bank.  
Whenever the user logs in, a series of coin tosses between user and
bank are used to generate a random $n$ digit substring of the $N$ digits,
where $n$ is significantly smaller than $N$.  
The user is then required to reveal only those $n$ digits of the 
passkey, and is accepted if they agree with the bank's records. 
This has the virtue that neither an eavesdropper nor someone impersonating
the bank can obtain very much of the passkey during a small number of
logins; in particular, neither has much chance of subsequently 
successfully convincing the bank that they are the user.   
By changing the passkey at appropriate intervals, security can thus
be maintained. 

Coin tossing also raises interesting theoretical questions, in that
its relation to other primitives has not so far 
been resolved.  

Bit commitment is another well known cryptographic 
primitive of great theoretical and practical interest, 
also involving two mistrustful parties. 
In a bit commitment protocol, one party, Alice, supplies 
an encoded bit to another, Bob.  Alice tries to ensure that Bob cannot
decode the bit until she reveals further information, while 
convincing Bob that she was genuinely committed all along. 
That is, Bob must be convinced that the protocol does not 
allow two different decodings of the bit which leave Alice 
free to reveal either $0$ or $1$, as she wishes. 

It is well known that secure coin tossing can easily be 
implemented given a secure bit commitment protocol.\refs{e.g. \kilian}   
Alice commits a random bit to Bob, who 
makes a random guess at it.  Alice then unveils the bit, and the parties 
generate (say) a $0$ if Bob's guess was correct, and a $1$ 
otherwise.   

This raises the question of whether the reverse is possible. 
As is by now well understood, quantum information has very different
properties from classical information.  In particular, some important
cryptographic tasks can implemented securely using quantum
information, but not using classical information.\refs{\wiesner,
\BBeightyfour} Thus, the question of the relation of coin tossing and
bit commitment subdivides into at least two independently interesting
questions: whether secure bit commitment can be built on top of secure
coin tossing using classical or quantum information exchanges.  Both
questions appear to have remained open to date (see
e.g. Ref. \refs{\mayersprl}).  My impression, for what it is worth, is
that most experts' best guess would have been that in the classical
case the reduction was probably impossible, while the quantum case was
regarded as anyone's guess.

There are many forms of security, of which the strongest and 
most interesting is {\it unconditional security}: an 
unconditionally secure protocol relies only on the known
laws of physics to ensure that the probability of successful
cheating by either party can be made arbitrarily small. 
Under standard non-relativistic 
cryptographic assumptions, unconditionally secure 
quantum bit commitment is impossible.\refs{\mayersprl, 
\mayerstrouble, \mayersone, \lochauprl, \lochau}  
We follow general usage in referring
to this result as the Mayers-Lo-Chau no-go theorem or 
MLC theorem.  

Unconditionally secure ideal coin tossing --- that is, coin 
tossing with probabilities precisely one half --- has also been
shown to be impossible by Lo and Chau.\refs{\lochau}  
However, it is not known whether non-ideal coin tossing, in 
which the probabilities are 
bounded by $(\half \pm \epsilon )$, and $\epsilon$ can be made
arbitrarily small, can be implemented with unconditional security 
in quantum theory.  Clearly, if it can be shown that 
quantum bit commitment can be built on top 
of coin tossing, then unconditionally secure quantum coin tossing
must be impossible.  Conversely, if it can be shown that quantum bit
commitment cannot be built on top of coin tossing, we have 
no conclusive argument showing that an unconditionally secure quantum
coin tossing protocol cannot be found. 
Such a protocol would be very useful. 

The main result of this paper resolves the relation between 
the two protocols by showing that secure bit commitment cannot be 
built on top of secure coin tossing in any finite classical or 
quantum protocol.  Though this result applies
to standard non-relativistic cryptography, its proof is 
inspired by considering cryptography in the context of 
relativity.  

The standard non-relativistic cryptographic scenario for two
mistrustful parties is as follows.  A and B each control a
laboratory, which includes sending and receiving equipment, 
measuring devices and classical and perhaps quantum computers. 
The laboratories are separated and 
generally assumed to be small.  A and B
have faith in the integrity of their own equipment, but trust
nothing whatsoever outside their laboratories.  In particular, 
neither of them has any way of ensuring that a message sent 
by the other was sent a certain time before receipt, and so 
an effectively simultaneous exchange of messages cannot be 
arranged.  A standard cryptographic protocol thus prescribes
a sequential exchange of messages between A and B, in which message
$i+1$ is not sent until the sender has received message
$i$.  

We will also need to consider an alternative cryptographic scenario
in which special relativity plays a r\^ole.  Alice and Bob agree on a 
frame and global coordinates, and on the location of two sites 
$\ul{x}_1, \ul{x}_2$ whose neighbourhoods Bob may control at all 
times.  Alice is not allowed within a distance $\epsilon$ of 
either point at any time.  
Alice is, however, required to erect laboratories within a 
distance $\delta$ of the sites, where 
$ \Delta x = | \ul{x}_1 - \ul{x}_2 | \gg \delta > \epsilon$.  
The precise location of Alice's laboratories need not be 
disclosed to Bob: he need only test that signals sent out from either of
his laboratories receive a response within time $2 \delta$.  
Bob could, for example, build laboratories of radius $\epsilon$
around each of the $\ul{x}_i$, but the precise location of 
his laboratories need not be known to Alice. 
She need only test that any signal broadcast from one of her
laboratories receives a reply within time $2 \delta$, whenever
her laboratory is in the prescribed region. 
Let the laboratories near $\ul{x}_i$ be $A_i$ and $B_i$, for $i = 1$ or $2$. 
We assume that the $A_i$ collaborate with complete mutual
trust and with prearranged agreement, and identify them
together simply as Alice; similarly the $B_i$ are identified as Bob.  
From the point of view of cryptographic analysis, any protocol in
this scenario may be considered as an two-party cryptographic 
protocol.  The only unusual cryptographic feature is that the 
parties each occupy disconnected laboratories, and even this is inessential: 
A and B could equally well occupy laboratories that are connected,
long, thin, and adjacent on their longer side.  The crucial 
difference from standard analyses is that the relativistic
signalling constraints which this situation imposes are taken into account.

There is a very simple unconditionally secure classical protocol 
for ideal coin-tossing under this circumstances.  At a pre-arranged time $t$, 
$A_1$ generates a random bit and sends it to $B_1$; at the same 
time, $B_2$ generates a random bit and sends it to $A_2$.  More
precisely, since the time of sending cannot be checked directly,
the bits are sent at or after time $t$, and so as to arrive before 
time $t+ \delta$ in each case.  The $A_i$ and $B_i$ then compare 
the bits they sent and received --- which, of course, involves a 
delay of order $\Delta x$.  If the two bits are equal, the 
protocol generates a $0$; if unequal, a $1$.  
The separation of the laboratories means that each party can
be confident that the other's bit was sent in ignorance of
their own.  Each party can hence be confident of 
the randomness of the generated bit.  
Quantum attacks clearly do not affect the 
protocol's security, since it relies only on the causal relations
of special relativity, so that the protocol also 
defines an unconditionally secure quantum coin-tossing protocol. 

On the other hand, a simple application of the Mayers-Lo-Chau 
argument\refs{\mayersprl, \lochauprl} shows
that no classical or quantum bit commitment protocol
that uses a finite sequence of messages can be permanently
unconditionally secure in this scenario.  By permanent security, 
we mean here that after the protocol is concluded, it remains
indefinitely impossible for Alice or Bob to cheat, no matter
how much information is transferred between the $A_i$ or 
between the $B_i$.  
Clearly, this cannot be attained by a finite protocol, since 
after the protocol is concluded, all data that the $A_i$ 
hold or receive can, after a finite time interval, be transferred 
to one representative, say $A_1$; similarly the $B_i$ can transfer
all their data to $B_1$.  At this point, the situation is identical 
to that after the implementation of a quantum bit commitment 
protocol in the standard scenario.  

To see this, note
that the $A_i$ and $B_i$ can carry out every step in any relativistic 
protocol at the quantum level.  In particular, any random choices
required by the protocol can be kept at the quantum level by 
entangling suitably chosen "quantum dice" --- ancillary systems
in a state $\sum_i p_i \ket{i}$ --- with the transmitted states via
a quantum computer.  If both $A_i$ (or both $B_i$) are required to
make the same random choice at spacelike separated points, they
can do so, using previously constructed shared random dice, 
with states of the form $\sum_i p_i \ket{i}_1 \ket{i}_2$, where
the $\ket{~}_i$ states are under the control of the $i$-th party. 
After the end of the finite protocol, all the quantum information
held by $A_2$ can be given to $A_1$, and all the quantum information held
by $B_2$ can be given to $B_1$.  We then have a situation in which $A_1$
and $B_1$ share some pure state $\ket{\psi}$ lying in the tensor product
Hilbert space ${\cal H}_A \otimes {\cal H}_B$, where 
${\cal H}_A$ and ${\cal H}_B$ describe the degrees 
of freedom under the control of $A_1$ and $B_1$ respectively. 
This is precisely the situation analysed by Mayers, Lo and Chau, 
and their theorem applies: either $B_1$ can cheat by 
distinguishing the commitments of $0$ and $1$ with non-zero 
probability before revelation, or else
$A_1$ can follow the protocol for committing a $0$ and then cheat
to reveal a $1$ with non-zero probability, and the two cheating
probabilities cannot simultaneously be 
small.\refs{\mayerstrouble,\lochau}

It follows that coin-tossing is a weaker primitive than bit 
commitment in the relativistic scenario outlined.  Perhaps
more surprisingly, it follows also that coin-tossing is weaker
under standard cryptographic assumptions.  For suppose there
were a finite standard bit commitment protocol which was provably 
secure modulo the security of a coin-tossing black box. 
That is, Alice and Bob have some trusted way of generating 
random bits between them, and build up a secure bit commitment
protocol by a finite sequence of classical or quantum communications
interspersed with finitely many invocations of the random bit 
generation.  
Any such protocol could be transferred to the relativistic 
scenario, by replacing the classical and quantum communications 
between $A$ and $B$ by identical communications between 
$A_1$ and $B_2$, and replacing the coin-tossing black box 
by implementations of the above secure quantum coin-tossing protocol
involving the $A_i$ and $B_i$.  
So long as the messages are carried out in the same sequence, the
bit commitment protocol would necessarily remain secure in the
relativistic scenario.  But we have seen that no secure finite
bit commitment protocol exists in this scenario.  Hence the 
initial assumption must be impossible: there is no finite 
permanently secure standard bit 
commitment protocol built on a secure coin-tossing
black box. 

It has recently been shown that 
unconditionally secure bit commitment protocols based on an
indefinite exchange of messages do exist in the relativistic 
scenario.\refs{\akbcrel,\akbcreltwo}  
We have seen here another 
intriguing aspect of the interplay between relativity and 
information: relativistic cryptography appears to provide
not only new practically useful protocols --- of which the 
coin tossing protocol above is an example --- but also a useful 
perspective on standard cryptographic relations.  
It should be noted, however, that while relativistic considerations
motivate the proof, they are not essential in establishing the 
result.  The only essential
relativistic ingredient we have used is the guarantee of a time delay in 
communicating between certain representatives of A 
and B, and the proof can be recast abstractly using only this
property. 

Finally, it is worth remarking that
cryptographic tasks give a way of calibrating the 
properties of information in {\it any} physical theory, correct
or not, by asking whether or not any given task can be securely implemented.
The fact that coin tossing is strictly weaker 
than bit commitment means they define distinct
calibrations of physical information.  
It would be interesting, and perhaps theoretically useful, to 
use the hierarchy of cryptographic protocols as a way of 
isolating different properties of information which can
be realised in different physical models. 

\leftline{\bf Acknowledgements}  
I am very grateful to
Claude Cr\'epeau, Hoi-Kwong Lo, David Mermin, Asher Peres,
Louis Salvail and, especially,
Peter Goddard and Sandu Popescu for very helpful comments, 
and to the Royal Society for financial support.

\listrefs

\end